\documentclass{emulateapj}
\usepackage{natbib}
\usepackage{multirow}
\usepackage{graphicx}
\shorttitle{Molecular Lines in Arp 220}
\shortauthors{Zschaechner, et al.}

\begin{document}

\title{High Resolution Observations of Molecular Lines in Arp 220: Kinematics, Morphology, and Limits on the Applicability of the Ammonia Thermometer} 
\author{\sc Laura K. Zschaechner\altaffilmark{1}, J\"{u}rgen Ott\altaffilmark{2}, Fabian Walter\altaffilmark{1}, David S. Meier\altaffilmark{3}, Emmanuel Momjian\altaffilmark{2}, and Nick Scoville\altaffilmark{4}}
\altaffiltext{1}{Max Planck Institute for Astronomy - K\"{o}nigstuhl 17, 69117 Heidelberg - Germany; zschaechner@mpia.de}
\altaffiltext{2}{National Radio Astronomy Observatory - P.O. Box O, 1003 Lopezville Road, Socorro, NM 87801, USA}
\altaffiltext{3}{Department of Physics, New Mexico Institute of Mining and Technology, 801 Leroy Place, Socorro, NM 87801, USA}
\altaffiltext{4}{California Institute of Technology, MC 249-17, 1200 East California Boulevard, Pasadena, CA 91125, USA}

\slugcomment{Accepted to The Astrophysical Journal on September 13, 2016}
\begin{abstract}

\par
      We observe Arp 220, the nearest Ultra-Luminous Infrared Galaxy (ULIRG), over 4 GHz in the K and Ka bands.   We provide constraints for the kinematics, morphology, and identify molecular species on scales resolving both nuclei (0.6" or 230 pc).  We detect multiple molecular species, including hydroxyl (OH $^{2}\Pi_{3/2} J=9/2 F=4-4; 5-5$) in both cores.  We tentatively detect H$_{2}$O(6$_{15}$-5$_{23}$) at $\sim$21.84 GHz in both nuclei, indicating the likely presence of maser emission.  The observed frequency range also contains metastable
ammonia transitions from (J,K) = (1,1) to (5,5), as well as the (9,9) inversion line, which, together are a well-known thermometer of dense molecular gas. Furthermore, the non-metastable (4,2) and (10,9) and possibly the (3,1) lines are also detected. We apply a standard temperature analysis to Arp 220. However, the analysis is complicated in that standard LTE assumptions do not hold.  There are indications that a substantial fraction of ammonia could be in the non-metastable transitions as opposed to only the metastable ones.  Thus, the non-metastable transitions could be essential to constraining the temperature.   We compare all of these data to ALMA observations of this source, confirming the outflow previously observed by other tracers in both nuclei.

\end{abstract}


\section{Introduction}\label{introduction}

\par
   Ultra-Luminous Infrared Galaxies (ULIRGs) exhibit abnormally high far infrared luminosities ($L_{IR}$ 8-1000 $\mu$m) of $>10^{12}L_{\odot}$ (\citealt{1987ApJ...320..238S}, \citealt{1996ARA&A..34..749S}). They are crucial to our understanding of how SF evolves (e.g.\ \citealt{2010MNRAS.405..219B}).  At a distance of 78 Mpc, Arp 220 is the nearest ULIRG, providing a unique opportunity to study these systems at high (380 pc arcsec$^{-1}$) resolution. Given Arp 220's location, it should not be surprising that it has drawn substantial interest over the years. Numerous observations at a variety of wavelengths have been obtained, as well as extensive simulations and empirical modeling.  Detections of molecular species include CO, originally detected by \citet{1986ApJ...311L..47S}, formaldehyde (H$_{2}$CO) detected by \citet{2004ApJS..154..541A} and multiple species by \citet{2008AJ....136..389S}, some of which were never before seen in an extragalactic source.  \citet{2011ApJ...742...95O} also observed the Hydroxyl (OH) $^{2}\Pi_{3/2} J=9/2 F=4-4; 5-5$ doublet (rest frequencies at 23.818 GHz and 23.827 GHz) in addition to the previously detected  $^{2}\Pi_{3/2} J=5/2 F=2-2$ \citep{2008AJ....136..389S}.  The former detection was confirmed by \citet{2013ApJ...779...33M} using GBT observations (which also included NH$_{3}$ metastable and non-metastable transitions).  \citet{2006ApJ...646L..49C} detect a potential H$_{2}$O(3$_{13}$-2$_{20}$) megamaser (rest frequency 183.310 GHz) with the IRAM 30-meter telescope, although no H$_{2}$O(6$_{15}$-5$_{23}$) maser was seen near 22.235 GHz (e.g.\ \citealt{1986A&A...155..193H}). We search for this maser in our data and report on it in $\S$~\ref{molecules}.  More generally, \citet{2011A&A...527A..36M} present an extended spectral survey of Arp 220 at higher frequencies, finding many molecular species.

\par
     Observations and interpretation of molecular lines are key to assessing the mechanisms driving SF in ULIRGs.  For example, ammonia (NH$_{3}$) traces the rotational temperature (T$_{rot}$), which can be used to approximate the kinetic temperature ($T_{kin}$).  Especially useful are the metastable (J=K) transitions. These metastable transitions have relative populations set by collisional processes, rendering them excellent indicators of $T_{kin}$.  Given this collisional dependency, however, the rotational temperature ($T_{rot}$) underestimates $T_{kin}$, especially for higher values of $T_{kin}$ (e.g.\ \citealt{1983A&A...122..164W}).  Thus, $T_{rot}$ should only be used as a lower limit.  Particularly relevant to this work is how near in frequency the inversion transitions are, allowing for simultaneous observations and thus the removal of complications resulting from inconsistent observing conditions and calibration (\citealt{1983A&A...122..164W}, \citealt{1988MNRAS.235..229D}). 

\par
   The initial detection of NH$_{3}$ in Arp 220 presented in \citet{2005PASJ...57L..29T} using single-dish Nobeyama data included the J=K (1,1) -- (4,4) transitions. Using the Australia Telescope Compact Array (ATCA) and Robert C. Byrd Green Bank Telescope (GBT) observations, \citet{2011ApJ...742...95O} detect the (5,5) and (6,6) inversion transitions.  \citet{2013ApJ...779...33M} present observations of (1,1)--(9,9) metastable transitions in Arp 220 with the GBT, and detect all but the (9,9) transition.  \citet{2013ApJ...779...33M}  also detect the (10,9) transition.  The two nuclei were not resolved in any of these observations. Higher resolution observations should reveal differences between the two nuclei, and stronger detections in cases where emission was diluted due to larger beam sizes.  

\par
   Radial motions associated with inflows or outflows, controlling both fueling and feedback, have substantial impact on SF on both local and global scales. Distinguishing regions (e.g.\ main disk, central regions, outskirts, shocked regions, or tidal features) in which certain molecules/transitions are found is necessary to discern when and where SF occurs.  Additionally, understanding the kinematics and morphology aids in assessing optical depth effects, which could potentially affect line ratios.  Arp 220 is already known to have an outflow (e.g.\ \citealt{2009ApJ...700L.104S}), which we explore further.

\par
    We utilize the Karl G. Jansky Very Large Array (VLA), with its newly available frequencies and extended bandwidths, to observe Arp 220.  With the VLA, it is possible to resolve the two nuclei and observe the higher $NH_{3}$ (9,9) transition ($\sim$850 K above ground) in addition to those observed in \citet{2011ApJ...742...95O}.  Furthermore, the observations presented here provide additional constraints on current models of the continuum such as those presented in \citet{2007A&A...468L..57D} and \citet{2015ApJ...799...10B}.  Complementary to VLA data are those taken of this source with ALMA (e.g.\ \citealt{2015ApJ...800...70S}, \citealt{2015ApJ...806...17R}).  We use both sets of observations to provide a more complete picture of the central regions of Arp 220 -- in particular the outflowing material.

\section{Observations and Data Reduction}\label{observations}

\par
   Observations were carried out using the VLA between March 2011 and June 2012 with the K-band (18-26.5 GHz) and Ka-band (26.5-40 GHz) receivers. An observational summary is given in Table~\ref{tbl_1}.

\begin{deluxetable}{lr}
\tabletypesize{\scriptsize}
\tablecaption{Observational and Instrumental Parameters  \label{tbl_1}}
\tablewidth{0pt}
\tablehead
{
\colhead{Parameter} &
\colhead{Value}\\
}
\startdata
\phd Observation Dates$-$K Band &2011 Mar 05\\
\phd &2011 Apr 17\\
\phd Observation Dates$-$Ka Band &2011 Mar 16\tablenotemark{a}\\
\phd &2012 May 25\\
\phd &2012 Jun 01\\
\phd &2012 Jun 02\\

\phd On-Source Time K Band (hours)&3.6\\
\phd On-Source Time Ka Band (hours)&3.2\\

\phd Pointing Center &15h 34m 57.225s\\
\phd &+23d 30m 11.57s\\ 

\phd Channel Spacing &250 kHz\\
\phd Spatial Resolution &0.6"\tablenotemark{b}\\
\phd RMS Noise (K Band 21.1--22.0 GHz) &0.33 mJy bm$^{-1}$\\
\phd RMS Noise (K Band 23.2--24.1 GHz) &0.42--0.62 mJy bm$^{-1}$\tablenotemark{c}\\
\phd RMS Noise (Ka Band 26.5--27.4 GHz) &0.70 mJy bm$^{-1}$\\
\phd RMS Noise (Ka Band 35.4--36.3 GHz) &0.70 mJy bm$^{-1}$\\

\enddata
\tablenotetext{a}{Data determined to be of poor quality and not used.}
\tablenotetext{b}{All cubes used for analysis of line data are smoothed to a circular 0.6" beam.} 
\tablenotetext{c}{ Because some spectral windows have half the integration time of others, the rms noise varies by $\sqrt{2}$.  Thus, the highest values of the rms noise are presented here.}
\end{deluxetable}

\par
   Data reduction was completed in Common Astronomy Software Applications (CASA, \citealt{2007ASPC..376..127M}) using standard spectral line methods.  For each receiver band two 1 GHz intermediate frequency channel (IF) pairs were utilized, both with right- and left- hand circular polarizations. From each 1 GHz IF pair eight 128 MHz subband pairs were correlated with 512 spectral channels per subband. The channel spacing was 250 kHz. The effective correlated frequency spans of each IF pair were 21.1-22.0 GHz, 23.2-24.1 GHz, 26.5-27.4 GHz, and 35.4-36.3 GHz.  The data are of good quality and minimal flagging was required.  The only noteworthy issue was that the amplitudes of individual spectral windows (spws) vary in a step-like fashion within the IF pair extending from 23.2 and 24.1 GHz in the scheduling block observed on 2011 April 17.  These variations are on the order of a few percent between individual spws.  Self-calibration slightly worsens the discontinuities, especially for spws containing large regions of absorption.  Thus, self-calibration is not used on the line data.

\par
   While the application of the self-calibrated solutions using the line-free channels on the spectral line data was disadvantageous as noted above, these issues do not affect most continuum bands.  Thus, we apply two iterations of phase-only self-calibration to the continuum alone (Figure~\ref{continuum_multiplot}) as it improves image quality without causing detrimental effects. 

\par  
   Total intensity (Stokes I) image cubes and maps were created in CASA using the task {\tt CLEAN}.  Natural weighting was used for all frequencies.  Frequencies at the high end of our spectral coverage yield a factor of $\sim$2 improvement in linear resolution over those at the lower end.  Thus to maintain consistency throughout the analysis, the spectral line cubes are convolved to the same beam size during imaging using {\tt CLEAN}.  Each cube consists of 3584$\times$250 kHz channels (2-3.4 km s$^{-1}$ depending on frequency).   Table~\ref{tbl_1} summarizes the properties of each image cube.

\section{The Data}\label{data}

\par

\begin{figure}
\begin{centering}
\includegraphics[width=70mm]{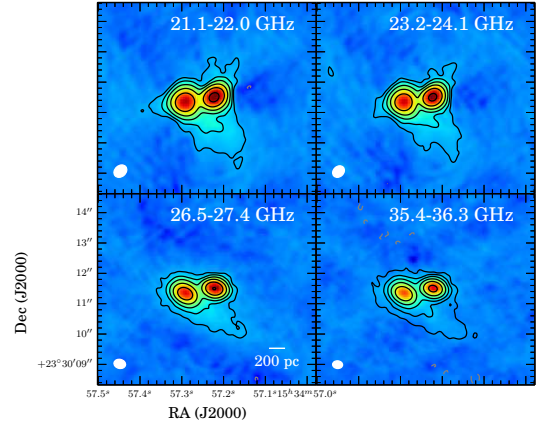}
\caption{\textit{Continuum images of both nuclei for each subband.  Note the south-western extension, which is real.  The northern and eastern extensions in the lowest contour in the top two panels are potentially artifacts as analogous features are not seen at other wavelengths. They also trace the artificial ``Y-shaped" VLA artifact pattern that is present at low but noticeable levels in the background.  This pattern is especially prominent in the top two panels. Contours correspond to 3$\sigma$ for each map (92, 87, 96, and 78 $\mu$Jy bm$^{-1}$ left to right, top to bottom) and increase by factors of 3.  3$\sigma$ negative contours are included in gray, but are only seen in the bottom-right panel. The beam is shown in the bottom-left corner of each panel.} \label{continuum_multiplot}}
\end{centering}
\end{figure}

\begin{figure*}
\begin{centering}
\includegraphics[width=150mm]{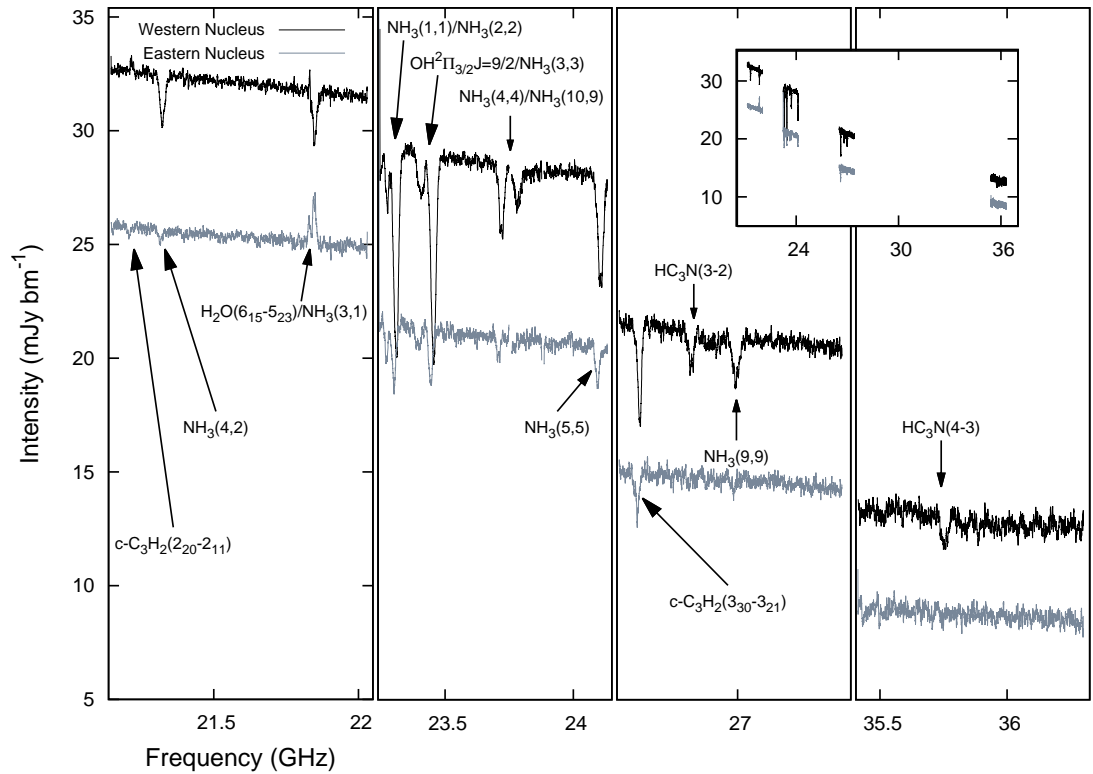}
\caption{\textit{The continuum and lines over all subbands averaging 10 contiguous channels.  Spectra from the peak of the western nucleus are shown in black while those from the eastern peak are shown in gray.  In the subband from 23 to 24 GHz, only the 2011 March 05 data are used with the exception of frequencies 23.750 and 23.878 GHz, which uses the 2011 April 17 data due to missing calibration information for the 2011 March 05 data in these spectral windows (note the gaps at the edges of this spectral window where troublesome channels were removed).  The data for this spectral window are scaled to match the overall slope of the 2011 March 05 data.  The inserted panel is a zoomed-out version to show the relative slope and frequencies of each subband.} \label{continuum_slope_compact_avg}}
\end{centering}
\end{figure*}

\par
   Stokes I continuum images clearly resolve both nuclei as well as a south-western extension (Figure~\ref{continuum_multiplot}).  Note that the northern extensions are possibly imaging artifacts.  The south-western extension is real and remarkably consistent with 6 and 33 GHz continuum observations presented by \citet{2015ApJ...799...10B}, although potentially affected by imaging artifacts as well.   We present these maps without smoothing to demonstrate that the beam sizes vary clearly with frequency, resulting in a factor of $\sim$4 improvement in beam area between the lowest and highest frequencies presented here.  This difference in beam size may affect the sensitivity to diffuse gas at higher frequencies, but it is also possible that the diffuse gas is only detected at lower energies regardless of beam area.  This consideration is particularly relevant to the south-western extension, which is seen clearly in the second contour in the top two panels, but is less prominent in the bottom two.

\par
   The full spectral extent of the data using beams smoothed to the same resolution at all frequencies is shown for the peaks of the two nuclei (Figure~\ref{continuum_slope_compact_avg}).  Note the smooth overall slope of the continuum.  The second (23.2-24.1 GHz) and third IF pairs (26.5-27.4GHz) contain the ammonia metastable transitions, (1,1)-(5,5) and (9,9).  Additional molecular species are detected and noted in Figure~\ref{continuum_slope_compact_avg} as well as in Table~\ref{tbl_3}.  A line complex is seen at $\sim$21.8 GHz in emission in the eastern nucleus, but in \textit{both} red-shifted emission and systemic/blue-shifted absorption in the western nucleus.  A line at $\sim$21.2 GHz is also seen purely in absorption in the westen nucleus, and in emission in the eastern nucleus.  We present possible identifications of these lines in $\S$~\ref{molecules}.

\par
These data represent the first resolved study of the two nuclei at these frequencies.  They improve upon the observations of \citet{2011ApJ...742...95O} and \citet{2013ApJ...779...33M} in both resolution and frequency coverage, allowing for simultaneous observations of the NH$_{3}$ (1,1)--(5,5) transitions in each nucleus for the first time.  These data, however, do not cover the (6,6) as in \citet{2011ApJ...742...95O} and \citet{2013ApJ...779...33M}, nor do they include the (7,7) and (8,8) transitions from \citet{2013ApJ...779...33M}.  Furthermore, the single dish data from \citet{2013ApJ...779...33M} allow for full flux recovery, which these data do not (although this is non-essential for the absorption-line studies presented here). The additional frequency coverage presented here allows for the detection of additional molecular species.  The increased resolution also enables detection of multiple molecular lines that were previously diluted by the larger beams.

\section{Results \& Discussion}\label{results}

\subsection{Molecular Species in Arp 220}\label{molecules}
\par
\begin{figure*}
\begin{centering}
\includegraphics[width=150mm]{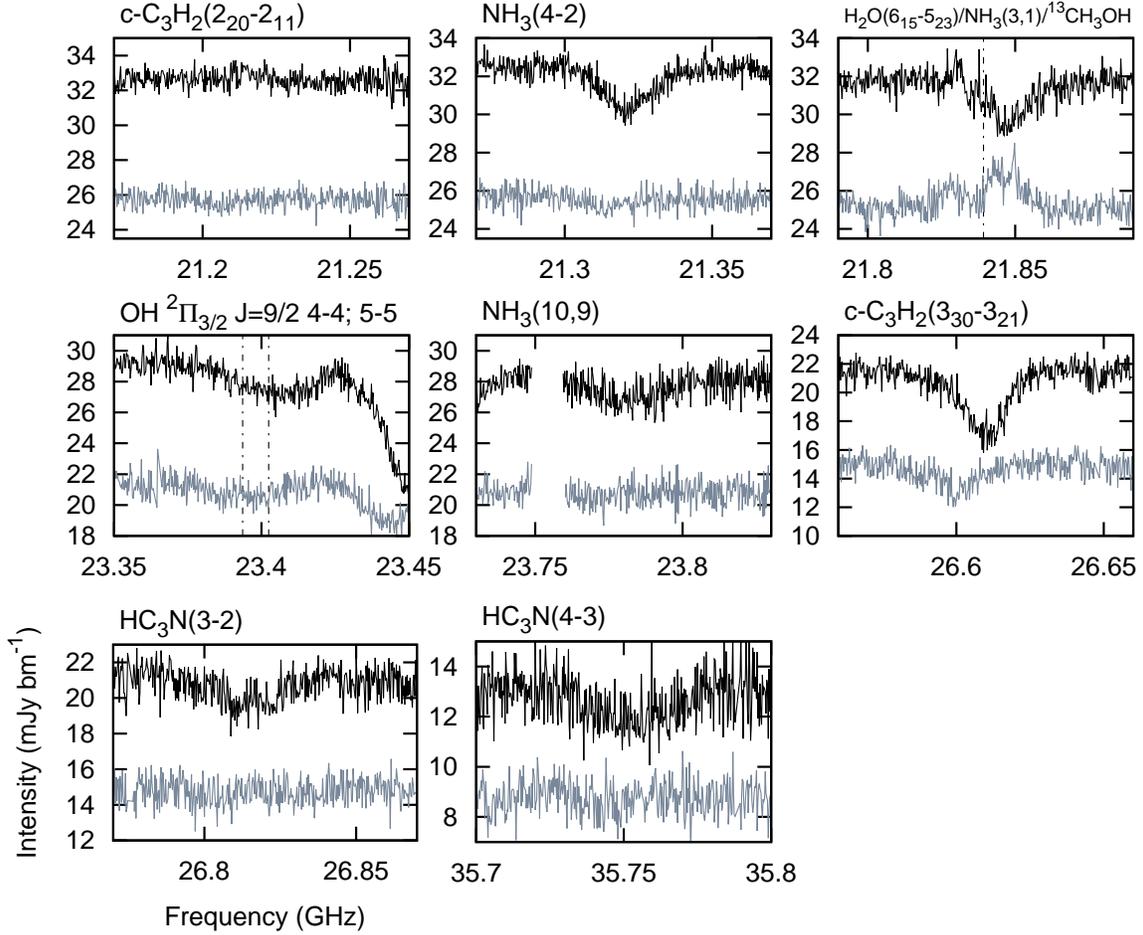}
\caption{\textit{Zoomed-in spectra from Figure~\ref{continuum_slope_compact_avg} for all lines except the ammonia metastable transitions.  As is the case for Figure~\ref{continuum_slope_compact_avg}, the western nucleus is shown in black and the eastern nucleus is shown in gray. The spectra in the upper right corner are likely a superposition of H$_{2}$O(6$_{15}$-5$_{23}$), NH$_{3}$(3,1), and possibly $^{13}$CH$_{3}$OH (see $\S$~\ref{molecules}).  Thus, all relevant transitions are displayed on top of the panel.  This spectrum is also shown in velocity units as the green line in Figure~\ref{spectrum_grid_west}. The systemic velocity of the entire system is indicated by dashed lines for H$_{2}$O(6$_{15}$-5$_{23}$) in the upper-right panel, and OH $^{2}\Pi_{3/2} J=9/2 F=4-4$ (lower frequency) and OH $^{2}\Pi_{3/2} J=9/2 F=5-5$ (higher frequency) in the left-most panel in the central row.  The gap in frequency coverage in the panel showing NH$_{3}$(10,9) is due to artefacts that have been removed.  The frequencies given on the x-axis are the observed frequencies.   Some lines are prominent in the western nucleus, yet faint or absent in the eastern nucleus (e.g.\ NH$_{3}$(4,2), NH$_{3}$(10,9), either HC$_{3}$N transition).   \  The dip seen at $\sim$23.45 GHz is (left-most panel, center row) NH$_{3}$(3,3). } \label{continuum_lines}}
\end{centering}
\end{figure*}

    Several molecular species are listed in Table~\ref{tbl_3} and shown in Figure~\ref{continuum_lines}, with the exception of the NH$_{3}$ metastable transitions, which we present in $\S$~\ref{temperature}.  Unless otherwise stated, identifications are based on the Splatalogue database \citep{2010AAS...21547905R}, and are the most likely candidates at those frequencies for this environment. 


\par
    We spatially resolve the OH $^{2}\Pi_{3/2} J=9/2 F=4-4; 5-5$ doublet at $\sim$23.4 GHz (originally detected at lower resolution by \citet{2011ApJ...742...95O} and confirmed by \citet{2013ApJ...779...33M}).  The doublet is detected in both nuclei. The two separate lines as observed are blended, forming a single, broad component.

\subsubsection{A Potential Detection of H$_{2}$O(6$_{15}$-5$_{23}$) in Arp 220}
\par
    Of particular interest is the detection at $\sim$21.84 GHz (top, right panel of Figure~\ref{continuum_lines}).  This complex is spatially-unresolved in the individual nuclei and exhibits both emission and absorption in the western nucleus, and emission in the eastern nucleus (with the possibility of absorption superimposed on the eastern nucleus).  The most likely candidates for this feature are NH$_{3}$(3,1) (rest frequency 22.234 GHz), H$_{2}$O(6$_{15}$-5$_{23}$) (rest frequency 22.235 GHz), and $^{13}$CH$_{3}$OH (rest frequency 22.239 GHz).  It is probable that there are at least two species present in the complex.  \citealt{2011A&A...527A..36M} detect multiple CH$_{3}$OH lines in Arp 220 between 210 GHz and 240 GHz.  These detections, however, are rather weak.  Thus, it is unlikely that $^{13}$CH$_{3}$OH could be responsible for a majority of the emission in the $\sim$21.84 GHz complex. Based on a previous detection by \citet{2006ApJ...646L..49C} H$_{2}$O(3$_{13}$-2$_{20}$) was thought to be a megamaser (rest frequency 183.310 GHz) centered at $\sim$5400 km s$^{-1}$ and having a velocity width of 350 km s$^{-1}$.  We conclude that most of the emission is likely due to H$_{2}$O(6$_{15}$-5$_{23}$), with NH$_{3}$(3,1) seen in absorption.

\par
   One could, in theory, estimate the NH$_{3}$(3,1) contribution based on the population of the other non-metastable states.  Nonetheless, given the unconstrained behavior seen in the metastable transitions (see $\S$~\ref{temperature}), only a very rough estimate can be made.  A RADEX \citep{2007A&A...468..627V} calculation assuming T$_{k}$$\sim$150 K and n=10$^{5-6}$ cm$^{-2}$ yields an intensity of the NH$_{3}$(3,1) transition approximately 2.5 times that of NH$_{3}$(4,2).  This ratio decreases almost linearly with T$_{k}$ and is close to 2 at 300 K.  The measured optical depth of the NH$_{3}$(4,2) transition in the western nucleus is comparable to the absorption in the H$_{2}$O(6$_{15}$-5$_{23}$)/NH$_{3}$(3,1) complex in that nucleus (although the latter shows some additional structure).  One possibility is that T$_{k}$ is substantially higher than indicated by data presented in $\S$~\ref{temperature} and in the literature.  Alternatively, it may be that the non-metastable transitions have departures from theory similar to those of the metastable transitions.  

\par
   However, it is important to keep the following in mind:  Given that NH$_{3}$(4,2) and NH$_{3}$(10,9) transitions do not prominently appear in absorption in the eastern nucleus, it may be that NH$_{3}$(3,1) absorption is not the source of the dip in the emission spectrum of the eastern nucleus.  There may be an additional molecular species present (e.g.\ $^{13}$CH$_{3}$OH), contributing to the absorption/emission in both nuclei. Alternatively, the dip may not be due to absorption at all, and may instead be a double-peaked kinematic structure related to disk rotation (consistent with \citealt{2016arXiv160509381S}).  The final molecular decomposition of this feature remains uncertain.  \citet{2015ApJ...799...10B} show that the star formation rate surface density (which should be correlated with the H$_{2}$O(6$_{15}$-5$_{23}$) maser luminosity) of the western nucleus is 2.5 times that of the eastern nucleus.  We multiply the H$_{2}$O(6$_{15}$-5$_{23}$) maser flux in the eastern nucleus by this factor.  We assume that the NH$_{3}$(3,1) is the dominant source of absorption in the western nucleus.  Based on these two assumptions, we find that the strength of the NH$_{3}$(3,1) line should be approximately twice of what is observed in that nucleus.  The resulting optical depth of the NH$_{3}$(3,1) transition is approximately $\tau$=0.06.  However, the RADEX calculations indicate that the optical depth of the NH$_{3}$(3,1) should be approximately twice that of the NH$_{3}$(4,2).  Given the observed $\tau$=0.07 for NH$_{3}$(4,2) in the western nucleus, the corresponding value for NH$_{3}$(3,1) should be $\tau$=0.14 or higher.  This discrepancy indicates that contamination from H$_{2}$O(6$_{15}$-5$_{23}$) maser emission may negate even more of the NH$_{3}$(3,1) absorption in the western nucleus than what the spectra at 22 GHz alone would indicate.

\par
    If we integrate the H$_{2}$O(6$_{15}$-5$_{23}$) emission over the eastern nucleus only, we obtain a luminosity of approximately 1.5$\times$10$^{8}$ K km s$^{-1}$ pc$^{2}$ over a velocity width of $\sim$180 km s$^{-1}$.  The strength of this detection is comparable to the 183 GHz emission detected by \citet{2006ApJ...646L..49C} of 2.5$\times$10$^{8}$ K km s$^{-1}$ pc$^{2}$ over 350 km s$^{-1}$.  (The 22 GHz emission only considers one nucleus -- the 22 GHz flux could increase by a factor of 2--2.5 when considering both nuclei.)  

\par
L$_{H_{2}O}$/L$_{FIR}$ is typically on the order of 10$^{-9}$ \citep{2005A&A...436...75H}.  Using methods presented in \citet{2005A&A...436...75H}, if we assume the factor of two higher luminosity derived for the observed H$_{2}$O(6$_{15}$-5$_{23}$) via the RADEX calculation, a factor of 3.5 increase when including both nuclei (based on the luminosity ratio of the two), and adopt L$_{FIR}$=1.4$\times$10$^{12}$ L$_{\odot}$ from \citet{2003AJ....126.1607S}, we find L$_{H_{2}O}$ $\sim$ 200 L$_{\odot}$ and L$_{H_{2}O}$/L$_{FIR}$$\sim$10$^{-10}$ for Arp 220.  This value is approximately an order of magnitude lower than what is typically seen in other galaxies with H$_{2}$O(6$_{15}$-5$_{23}$) detections.  However, it is possible that there could be contamination from NH$_{3}$ (3,1) absorption in the eastern nucleus as well.  This potential contamination could result in a value up to four times higher for the H$_{2}$O(6$_{15}$-5$_{23}$) luminosity.  Such a scenario would render L$_{H_{2}O}$/L$_{FIR}$=10$^{-9}$, which is consistent with the sample observed in \citet{2005A&A...436...75H}.

\par
A detection of a H$_{2}$O(6$_{15}$-5$_{23}$) megamaser in Arp 220 is interesting since it is one of the few in a ULIRG reported to date.  Aside from this detection, \citet{2016ApJ...816...55W} detect a $\sim$1600 L$_{\odot}$ H$_{2}$O(6$_{15}$-5$_{23}$) megamaser in the ULIRG UGC 5101.  \citet{1984A&A...141L...1H}, \citet{2015ApJ...815..124H}, and \citet{2002PASJ...54L..27N} present H$_{2}$O(6$_{15}$-5$_{23}$) detections in NGC 6240 (LIRG) on the order of a few L$_{\odot}$, which is substantially lower than what we detect in Arp 220.  However, the L$_{FIR}$ of NGC 6240 is 10$^{11}$--10$^{12}$ L$_{\odot}$ -- up to an order of magnitude lower than that of Arp 220.

\par
It is important to emphasize that, in spite of the detection at 183 GHz, no emission has previously been seen near 22 GHz in Arp 220 - only an upper limit of 0.15 Jy set by \citet{1986A&A...155..193H} for both nuclei (unresolved).  However, neither the \citet{1986A&A...155..193H} nor the 183 GHz observations by \citet{2006ApJ...646L..49C} resolved the two nuclei.  After creating a zeroth-moment map including all the channels in which emission from this line is seen, we find approximately $-$60 Jy bm$^{-1}$ km s$^{-1}$ in the western nucleus and $+$60 Jy bm$^{-1}$ km s$^{-1}$ in the eastern nucleus (the two values are within 3 Jy bm$^{-1}$ km s$^{-1}$ of each other), thus canceling all but a negligible fraction of the total emission.  Thus, when the nuclei are blended together such as they are when observed at lower resolution, the result is a non-detection.

\begin{deluxetable*}{lcccccc}
\tabletypesize{\scriptsize}
\tablecaption{Molecular Species Excluding Metastable Ammona \label{tbl_3}}
\tablewidth{0pt}
\tablehead
{ 
\colhead{$\nu_{obs}$ (GHz)}&
\colhead{Molecular Species\tablenotemark{a}} &
\colhead{$\nu_{rest}$ (GHz)}&
\colhead{$\tau_{peak}$}&
\colhead{$v_{peak}$ (km s$^{-1}$)}&
\colhead{${\Delta}v_{1/2}$ (km s$^{-1}$)}&
\colhead{${\int}{\tau}dv$ (km s$^{-1}$)}\\
}
\startdata
\\
\phd Western nucleus&&&&&& \\
\\
\phd 21.22&c-C$_{3}$H$_{2}$(2$_{20}$-2$_{11}$)&21.58740& \tablenotemark{b} &&& \\ 
\phd 21.32&NH$_{3}$(4,2)& 21.70341 & 0.07$\pm$0.002 & $-$115$\pm$4 & 118$\pm$4 & 8.8$\pm$0.4\\

\phd 21.84&H$_{2}$O(6$_{15}$-5$_{23}$), NH$_{3}$(3,1)\tablenotemark{c} &22.23508, 22.23456 & &&&\\ 
\phd &$^{13}$CH$_{3}$OH &22.23992 & &&&\\ 
\phd 23.40&OH $^{2}\Pi_{3/2}$ J=9/2&23.8176153, 23.8266211 & 0.05$\pm$0.005 & $-$159$\pm$16 & 124$\pm$16 &6.6$\pm$1.1 \\
\phd 23.78&NH$_{3}$(10,9)\tablenotemark{d}& 24.20536 & 0.05$\pm$0.004 & $-$143$\pm$13 & 163$\pm$14  & 8.7$\pm$1.0\\
\phd 26.61&c-C$_{3}$H$_{2}$(3$_{30}$-3$_{21}$) & 27.08435 & 0.23$\pm$0.13 & $-$133$\pm$51 & 79$\pm$51 & 19.4$\pm$16.7\\
\phd 26.82&HC$_{3}$N(3-2)\tablenotemark{e}& 27.29290-27.29623 & 0.10$\pm$0.05 & $-$135$\pm$71 & 116$\pm$71 & 12.4$\pm$9.8\\
\phd 35.75&HC$_{3}$N(4-3)\tablenotemark{e}& 36.39-36.394 & 0.10$\pm$0.009&$-$126$\pm$10 & 100$\pm$10 & 10.7$\pm$1.4\\

\\
\phd Eastern nucleus&&&&&& \\
\phd 21.21&c-C$_{3}$H$_{2}$(2$_{20}$-2$_{11}$)& 21.58740 &  &  &  &\\ 
\phd 21.31&NH$_{3}$(4,2)& 21.70341 & 0.03$\pm$0.006 & 5$\pm$9 & 38$\pm$9 & 1.2$\pm$0.4 \\
\phd 21.84&H$_{2}$O(6$_{15}$-5$_{23}$), $^{13}$CH$_{3}$OH\tablenotemark{c}& 22.23508, 22.23992 &  &  &  &\\
\phd 23.39&OH $^{2}\Pi_{3/2}$ J=9/2&23.8176153, 23.8266211 & 0.03$\pm$0.005 & $-$30$\pm$27 & 150$\pm$27 & 4.8$\pm$1.2\\
\phd 26.60&c-C$_{3}$H$_{2}$(3$_{30}$-3$_{21}$) & 27.08435 & 0.11$\pm$0.10 & $-$20$\pm$107 & 99$\pm$107 & 11.6$\pm$15.1\\

\enddata
\tablenotetext{a}{Molecular species are identified using the Splatalogue database unless previously identified in the Arp 220 related literature.  Each line is among the brightest in its respective frequency band and is matched to the systemic velocity within the uncertainties.}
\tablenotetext{b}{The optical depth and kinematic parameters are not given in cases where the line is not cleanly fit by a gaussian due to additional structures present that cannot be unambiguously distinguished between kinematic and morphological features, as well as additional molecular species.}
\tablenotetext{c}{There are indications of multiple lines present.  A tentative ID of H$_{2}$O(6$_{15}$-5$_{23}$) is based on presence of H$_{2}$O(3$_{13}$-2$_{20}$) detected by \citet{2006ApJ...646L..49C} and the most likely molecules at this frequency in the Splatalogue database.  $^{13}$CH$_{3}$OH may also be present, although it would likely contribute less to the emission in this frequency range.}
\tablenotetext{d}{With the GBT, \citet{2013ApJ...779...33M} detect NH$_{3}$(10,9) at velocities consistent with the \textit{eastern nucleus, but not the western}.  Our data show indications of NH$_{3}$(10,9) absorption towards the eastern nucleus, but no unambiguous detection.  In the western nucleus the NH$_{3}$(10,9) absorption is clearly present.}
\tablenotetext{e}{Hyperfine structure lines are present, but unresolved.}
\end{deluxetable*}

\subsection{Temperature Analysis Using Metastable Ammonia}\label{temperature}

\subsubsection{Method}\label{method}
\par
\begin{figure}[!htb]
\begin{centering}
\includegraphics[width=100mm]{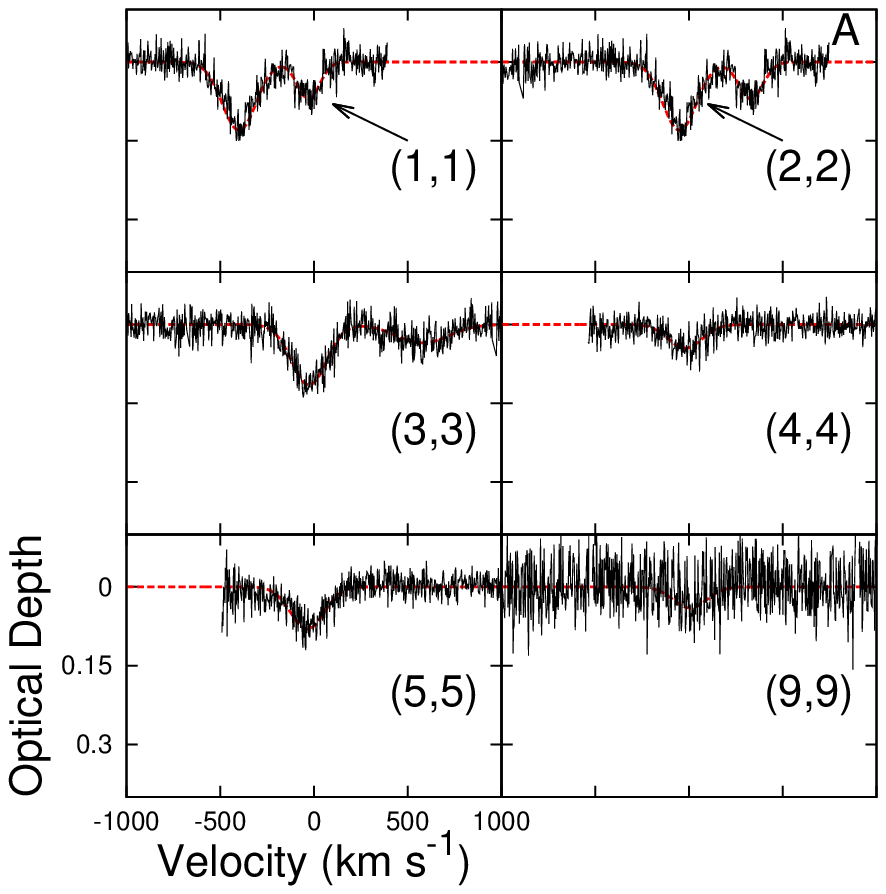}
\includegraphics[width=100mm]{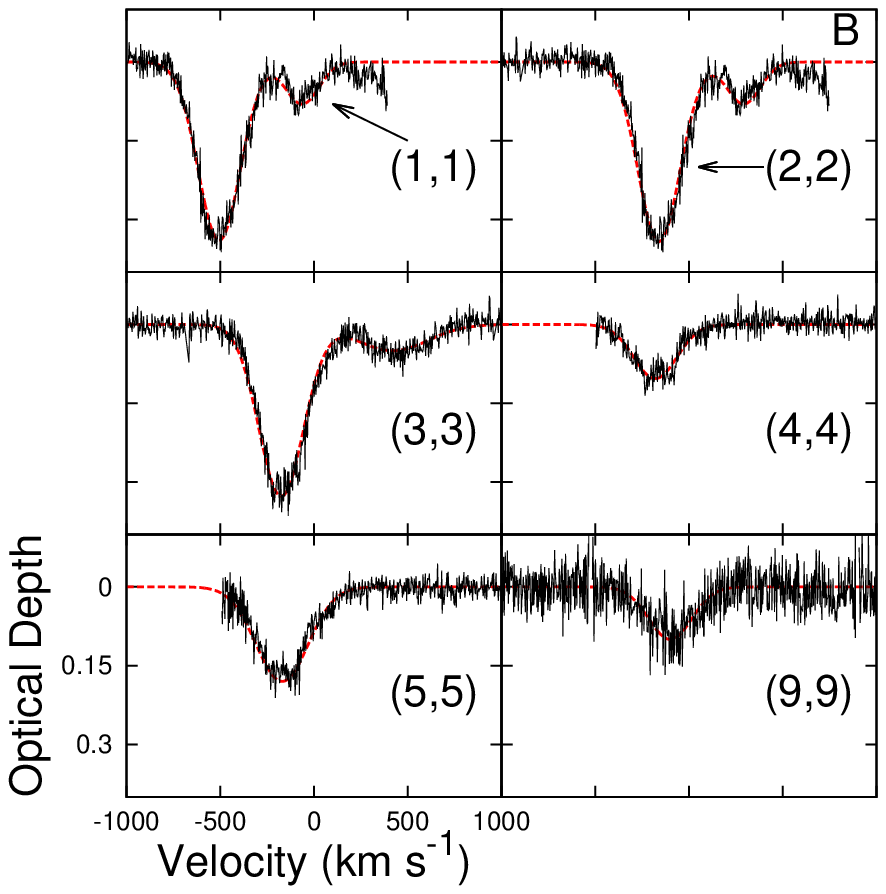}
\caption{\textit{Spectral fits for the metastable ammonia inversion transitions in the eastern nucleus (A), and the western nucleus (B).  The top two panels show both the (1,1) and (2,2) inversion transitions, each shifted accordingly to the systemic velocity for their respective fits.  The (9,9) inversion transition was problematic during fitting due to the increased rms noise at higher frequencies and intrinsically shallow depth of the line.  Finer structures may indeed be present in some transitions (e.g.\ the (4,4) and (5,5) transitions in the western nucleus), but we fit these as single gaussians for the purposes of our analysis.  These fits are used to determine the optical depth and line-width, which are then used in the temperature analysis presented graphically in Figure~\ref{boltzmann_combined}.} \label{spectra_west}}
\end{centering}
\end{figure}

     Here we detect the NH$_{3}$ (1,1) -- (5, 5) inversion lines in absorption (see Figure~\ref{spectra_west} and Table~\ref{tbl_2}).  Additionally, we detect the (9, 9) inversion line, which, in theory, substantially increases our leverage in determining $T_{rot}$ and $T_{kin}$.

\begin{figure*}[!htb]
\begin{centering}
\includegraphics[width=130mm]{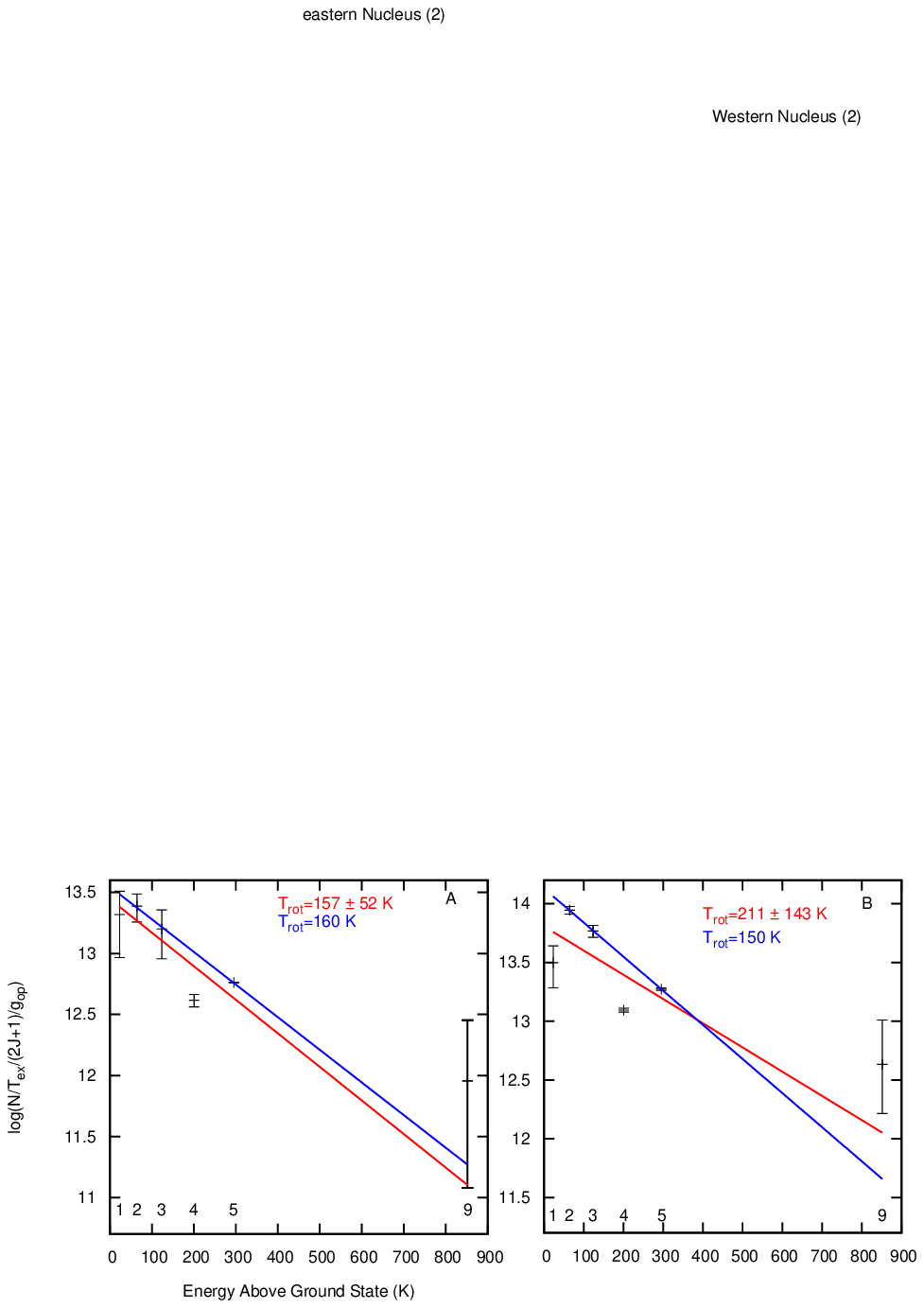}
\caption{\textit{Boltzmann plots of the metastable ammonia inversion transitions in the eastern nucleus (A), and the western nucleus (B).  Note the relatively low population of the (1,1) transition.  Also, the population of the (9,9) transition is relatively high in both nuclei, likely better fit by a two-temperature solution.  The (4,4) and (5,5) lines are also inconsistent with expectations relative to each other, while it is not clear which transition is problematic.  The red line fits the (1,1)--(5,5) transitions, while the blue line fits the (2,2), (3,3), and (5,5) transitions only in order to show how T$_{rot}$ changes when the obviously discrepant data points are discarded.  For the western nucleus, there  is a substantial difference in rotational temperatures determined with each.} \label{boltzmann_combined}}
\end{centering}
\end{figure*}

We use methods involving Boltzmann statistics such as those presented in \citet{2005PASJ...57L..29T}, \citet{2011ApJ...742...95O}, and \citet{2013ApJ...772..105M} to gauge $T_{rot}$, which can then be used to estimate $T_{kin}$.  We adopt the assumption that the relative populations of each rotational transition are related to $T_{rot}$ \citep{1983A&A...122..164W}.  We first determine a ratio of the column densities in the upper state ($N_{u}$) to excitation temperature $T_{ex}$.

\begin{equation}
\frac{N_{u}}{T_{ex}}=7.28\times10^{13}\frac{J(J+1)}{K^{2}\nu}\times\tau\times\Delta{v}_{1/2}
\end{equation} 
   
    where $\tau$ is the central optical depth, $\nu$ is the central frequency in GHz and $\Delta{v}_{1/2}$ is the FWHM in km s$^{-1}$.  This equation is derived from Equation 30 in \citet{2015PASP..127..266M} under the assumption of using the peak optical depths and FWHMs (otherwise the prefactor would be 7.74).  Furthermore, this prefactor differs from the often used 1.61$\times$10$^{14}$ (e.g.\ \citealt{1995A&A...294..667H}, \citealt{2011ApJ...742...95O}, \citealt{2013ApJ...772..105M}) by approximately a factor of two as it only includes one inversion level as opposed to two.   Taking into account the statistical weight of each transition and using a Boltzmann law we obtain:

\begin{equation}
\frac{N_{u'}/T'_{ex}}{N_{u}/T_{ex}}=\frac{g_{op}(J')}{g_{op}(J)}\frac{2J'+1}{2J+1}exp\bigg(\frac{-{\Delta}E}{T_{rot}}\bigg),
\end{equation}

  where $g_{op}$ is the statistical weighting factor of the line ($g_{op}=1$ for para and $g_{op}=2$ for ortho). The log of the weighted column densities is plotted against the energy of the upper level above the ground state in Kelvin.  A line (or in the case of clear departures from a gaussian, a multiplet) is fit to the data (Figure~\ref{boltzmann_combined}). In the end $T_{rot}=-log_{10}(e)/m$ where $m$ is the slope of the best fit line.  $T_{rot}$ can then be used to approximate $T_{kin}$.

\begin{deluxetable*}{lcccccc}
\tabletypesize{\scriptsize}
\tablecaption{Line Parameters - Metastable Ammonia \label{tbl_2}}
\tablewidth{0pt}
\tablehead
{
\colhead{Line} &
\colhead{$\nu_{rest}$ (GHz)}&
\colhead{$\nu_{obs}$ (GHz)}&
\colhead{$\tau_{peak}$}&
\colhead{$v_{peak}$ (km s$^{-1}$)}\tablenotemark{a}&
\colhead{${\Delta}v_{1/2}$ (km s$^{-1}$)}&
\colhead{${\int}{\tau}dv$ (km s$^{-1}$)}\\
}
\startdata
\\
\phd Western nucleus&&&&&& \\
\phd NH$_{3}$ (1,1)&23.69477&23.27&0.069$\pm$0.017\tablenotemark{b}&$-$63$\pm$26&210$\pm$62&15.4$\pm$6.0\\
\phd NH$_{3}$ (2,2)&23.72260&23.30&0.342$\pm$0.016&$-$150$\pm$6&264$\pm$15&95.9$\pm$7.2\\
\phd NH$_{3}$ (3,3)&23.87008&23.45&0.326$\pm$0.034&$-$163$\pm$11&286$\pm$25&99.2$\pm$11.6\\
\phd NH$_{3}$ (4,4)&24.13935&23.71&0.104$\pm$0.003&$-$182$\pm$3&269$\pm$8&29.7$\pm$1.1\\
\phd NH$_{3}$ (5,5)&24.53292&24.10&0.169$\pm$0.003&$-$165$\pm$3&322$\pm$6&58.1$\pm$1.5\\
\phd NH$_{3}$ (9,9)&27.47794&26.99&0.096$\pm$0.087&$-$100$\pm$120&271$\pm$282&27.8$\pm$38.5\tablenotemark{c}\\
\\
\phd Eastern nucleus&&&&&& \\
\phd NH$_{3}$ (1,1)&23.69477&23.27&0.067$\pm$0.024&$-$30$\pm$25&142$\pm$60&10.2$\pm$5.7\\
\phd NH$_{3}$ (2,2)&23.72260&23.30&0.115$\pm$0.019&$-$42$\pm$18&216$\pm$42&26.5$\pm$6.9\\
\phd NH$_{3}$ (3,3)&23.87008&23.45&0.115$\pm$0.032&$-$26$\pm$30&219$\pm$71&26.9$\pm$11.6\\
\phd NH$_{3}$ (4,4)&24.13935&23.71&0.045$\pm$0.003&$-$19$\pm$7&204$\pm$18&9.9$\pm$2.0\\
\phd NH$_{3}$ (5,5)&24.53292&24.10&0.071$\pm$0.004&$-$35$\pm$6&235$\pm$14&17.9$\pm$1.42\\
\phd NH$_{3}$ (9,9)&27.47794&26.99&0.026$\pm$0.08&5$\pm$396&216\tablenotemark{c}&5.9$\pm$17.7\\
\enddata
\tablenotetext{a}{Velocity centers are given with respect to the systemic velocity (optical, heliocentric) of Arp 220, 5434 km s$^{-1}$.}
\tablenotetext{b}{Due to the possibility of incomplete coverage of the continuum by molecular clouds or contamination by cooler gas, the true optical depths are likely somewhat higher than what we measure here.}
\tablenotetext{c}{The uncertainties for the (9,9) inversion transition are quite high for both nuclei.  Due to a low SNR, we are unable to fit the line-width of the (9,9) inversion transition in the eastern nucleus.  Thus, we force a value of 216 km s$^{-1}$ (consistent with the other line widths for the eastern nucleus) for the purpose of determining the optical depth.}
\end{deluxetable*}

\par
    It is immediately clear that these data are not described by a straight line, consistent with a single T$_{rot}$, \textit{which limits what can be gleaned.}  An assumption for this method is LTE, which the two nuclei do not fulfill.  For illustrative purposes, to compare with existing work, and to gauge the extent of the impact of non-metastable transitions, we present the most basic analysis steps.

\par
  The slope of Boltzmann diagrams (Figure~\ref{boltzmann_combined}) typically shallows for higher metastable transitions (e.g.\ \citealt{2013ApJ...772..105M}).  Thus, the seemingly overpopulated (9,9) transition is unsurprising.  Given that there are no additional higher-order ortho transitions observed that could aid in constraining a two-temperature model, we omit the (9,9) transition from the analysis. The (1,1) transition is underpopulated for both nuclei, which was also observed by \citet{2011ApJ...742...95O} and \citet{2013ApJ...779...33M}.   \citet{2011ApJ...742...95O} suggested cooler surrounding gas as an explanation.   \citet{2013ApJ...779...33M} suggest that the (2,2) transition is broadened and amplified by an additional velocity component, potentially associated with an outflow. However, the latter scenario does not account for the (5,5) transition being deeper in absorption than expected relative to the (4,4) transition, which is apparent in the lower-resolution data of \citet{2013ApJ...779...33M} and which we clearly see in both nuclei.   %

\par
  To illustrate how much the rotational temperature is affected by these departures from linearity, we make two estimates of the rotational temperature.  In the first we include the (1,1) through (5,5) transitions.  In the second, we omit the (1,1) and (4,4) transitions.  Thus, the second estimate excludes all transitions for which we are \textit{certain} do not conform to the expectations set forth by this method (Figure~\ref{boltzmann_combined}).  The differences between the two estimates gives an indication of the impact of anomalous behavior on determinations of the rotational temperature.

\par
Relevant to the break-down of this analysis in the case of Arp 220, two non-metastable lines are also detected at comparable levels (and possibly a third -- the (3,1) transition, although this line is likely blended).  Thus, the non-metastable transitions may be substantially influencing the derived excitation.  For example, the (2,1) and (3,1) states could be substantially populated which would affect the (1,1) population.  If the populations of the metastable (1,1) and the higher excitation (2,1) and (3,1) transitions are combined, then the expected population of the (1,1) state could be recovered.  Naturally, these principles would apply to the other metastable transitions. To assess the likelihood of this explanation, we examine the (3,1) and (4,2) transitions for the Western nucleus.  Using the same method as for the metastable transitions, if the (3,1) transition is included with the (1,1) transition, there is indeed a substantial effect - in fact the (3,1) state is more populated than the (1,1) transition, and the value for the (1,1) transition on the y-axis of Figure~\ref{boltzmann_combined} would increase to approximately 14.3 when they are combined (the new value is not plotted).  The same can be done for the (4,2) and (2,2), and a more modest increase is seen to approximately 14.1.  Note that \textit{the (1,1) transition is now higher than the (2,2) transition -- consistent with the assumptions used to derive $T_{rot}$}.  This is after adding only one nonmetastable transition to each metastable one, which is nowhere near a complete analysis. Although not all metastable transitions are included here, we can already see that they could potentially account for the break-down of the metastable temperature analysis in Arp 220.

\par
   We forgo further Large Velocity Gradient (LVG) and other analyses of these data since it is abundantly clear that even a basic analysis already results in a relatively poor fit to derive the rotational temperature.  Additional data on the non-metastable transitions are required to derive further constraints.

\par
   Consistent with the findings of \citet{2011ApJ...742...95O}, there is no clear indication of an ortho-para ratio far from unity.   Thus, we can assume an ortho-para ratio close to one in both nuclei, indicating warm conditions when the ammonia formed.

\subsection{An outflow demonstrated by a comparison with emission 
lines seen by ALMA}\label{outflow}

\begin{figure}[!htb]
\begin{centering}
\includegraphics[width=70mm]{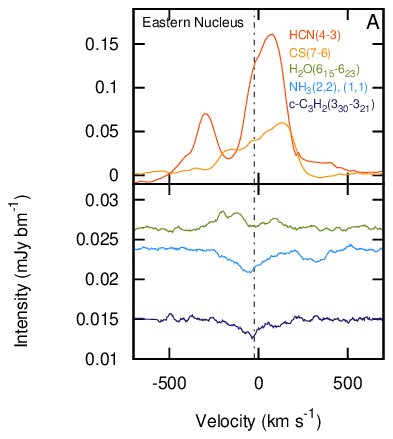}
\end{centering}
\end{figure}
\begin{figure}[!htb]
\begin{centering}
\includegraphics[width=70mm]{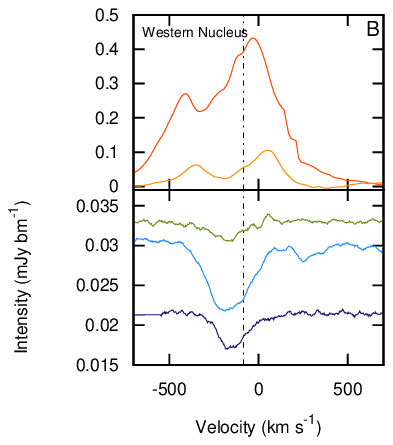}
\caption{\textit{We present representative molecular species from the sample of those detected in Arp 220.  The systemic velocity of the entire system (LSR (radio) 5434 km s$^{-1}$) is at zero.  The systemic values of each nucleus taken from \citet{2009ApJ...700L.104S} are $-$84 km s$^{-1}$ for the western and $-$24 km s$^{-1}$ for the eastern nucleus (also LSR (radio)).  The top panels show ALMA observations of HCN(4-3) (red) and CS (7-6) (orange) by \citet{2015ApJ...800...70S}.  The bottom panels show the H$_{2}$O(6$_{15}$-5$_{23}$)/NH$_{3}$(3,1) complex (green), NH$_{3}$(2,2) (light blue), and what we identify here as c-C$_{3}$H$_{2}$(3$_{30}$-3$_{21}$) (dark blue).  All lines belonging to the bottom panels are listed in Tables~\ref{tbl_3} or~\ref{tbl_2}. The NH$_{3}$(2,2) spectra also include the NH$_{3}$(1,1) inversion transition (the shallower line to the right side).  The self-absorption in the emission detected by ALMA is closely duplicated by the absorption lines seen with the VLA.  In particular, the absorption in the western nucleus matches the CS(7-6) profile almost exactly (vertical scaling factors aside).  The absorption appears to be blue-shifted with respect to the systemic velocities of both nuclei (dotted lines), indicating outflow is likely (more so for the western nucleus).} \label{spectrum_grid_west}}
\end{centering}
\end{figure}

\par
   We compare the absorption spectra we observe with the VLA with HCN (4-3) and CS (7-6) spectra observed with ALMA by \citet{2015ApJ...800...70S}.   The HCN (4-3) and CS (7-6) spectra show a decrease or dip in the spectrum of each nucleus (Figure~\ref{spectrum_grid_west}).  In the western nucleus, the decrease is clearly blue-shifted with respect to the systemic velocity of that nucleus.  The dip in the eastern nucleus is also blue-shifted, but to a lesser degree.  A double-peak is also seen in the eastern nucleus.   Considering also the CO(1-0) observations shown in Figure 3 of \citet{2016arXiv160509381S}, which reveal no analogous feature in the single-peaked CO (1-0) spectrum, it is clear that the decrease in the western nucleus is unlikely to be a kinematic feature.  Thus, this decrease in the western nucleus is almost certainly due to absorption.  The cause of the dip in the eastern nucleus is less clear as it is consistent with the two-peaked spectrum seen in the CO (1-0) at the farthest extent, also presented in \citet{2016arXiv160509381S}.  When the self-absorption in the ALMA data is compared to the absorption in the VLA data in the western nucleus, one can see that they are aligned with each other.    Assuming the LSR (radio) value of 5350 km s$^{-1}$ for the systemic velocity of the western nucleus \citep{2009ApJ...700L.104S}, the absorption spectra in that nucleus are blue-shifted to $\sim$80 km s$^{-1}$ at the peak and $\sim$400 km s$^{-1}$ at the farthest extent (with the exception of the HCN (4-3)).  Evidence for blue-shifted absorption is less clear in the eastern nucleus, but may be on the order of $\leq$30 km s$^{-1}$ (adopting a systemic velocity for that nucleus of 5410 km s$^{-1}$ based on \citealt{2009ApJ...700L.104S}).

\par
   The blue-shifted absorption indicates outflowing material, consistent with the picture presented in \citet{2009ApJ...700L.104S} and \citet{2015ApJ...800...25T}.  \citet{2001ApJ...560..168M} and \citet{2009ApJ...700L.104S} noted outflowing CO in absorption with a velocity on the order of $\sim$100 km s$^{-1}$. Additionally, \citet{2015ApJ...800...25T} observed a symmetric outflow seen in velocities up to $\pm$500 km s$^{-1}$ in the western nucleus, which included multiple molecular species (e.g.\ SiO(6,5), H$^{13}$CN(3,2)).  That we only see a blue-shift of only 400 km s$^{-1}$ in the western nucleus and even less in the eastern nucleus indicates that the outflow seen in absorption in our observations, is either impeded, driven by a different mechanism, and/or from emission confined to different radii.   

\par 
    CS(7-6) and the absorption features we observe with the VLA show greater similarities than with those of the HCN(4-3).  This indicates that the CS(7-6) and VLA features likely both originate from material at similar radii and share similar kinematics.  The comparatively blue-shifted (on the order of 100 km s$^{-1}$ or more) self-absorption in the HCN(4-3) line could indicate different kinematics, presence in different parts of the western nuclear region, or both.  If the HCN(4-3) is indeed present at larger radii, then the entirety of its emission is likely less affected by absorption against the continuum core.  Furthermore, the material most likely to be self-absorbed would be that directly between the continuum core and the observer -- having the largest line-of-sight velocities (consistent with the relative blue-shift).  This picture is consistent with figures in \citet{2015ApJ...800...70S} which show the HCN(4-3) line having a larger spatial extent than CS(7-6).  Furthermore, the kinematic axes of HCN(4-3) and CS(7-6) are offset by 45$^{\circ}$, indicating distinct morphologies, with the VLA absorption we present here conforming more to the latter.

\section{Summary}\label{summary}

\par
1)  We observe the two nuclei of Arp 220 at $\sim$0.6" resolution for selected frequencies between 21 and 37 GHz. This is the first study that resolves the two nuclei at these frequencies. We present continuum maps resolving the two nuclei and a south-western extension, as well as a full spectrum of the peak of each nucleus for these frequencies.

\par
2)  In both nuclei, we detect (OH) $^{2}\Pi_{3/2} J=9/2 F=4-4; 5-5$, also seen with lower spatial resolution by \citet{2011ApJ...742...95O} and \citet{2013ApJ...779...33M}.
\par
3)  We tentatively detect H$_{2}$O(6$_{15}$-5$_{23}$) at $\sim$21.84 GHz in the eastern nucleus. In the eastern nucleus, the apparent line is seen entirely in emission, although some possible absorption is evident.  In the western nucleus, the line is seen primarily in absorption, with a blue-shifted emission component.  We interpret this absorption as NH$_{3}$(3,1), which cancels most of the H$_{2}$O(6$_{15}$-5$_{23}$) in the western nucleus.  It is also possible that $^{13}$CH$_{3}$OH is present, although it would likely be weak compared to the other lines.   The relative contributions of the two nuclei are similar in magnitude, but opposite sign resulting in a net non-detection for unresolved observations, such as those presented by \citet{1986A&A...155..193H}.  When considering the H$_{2}$O(6$_{15}$-5$_{23}$) luminosity, we find a value of 10$^{-10}$--10$^{-9}$ for L$_{H_{2}O}$/L$_{FIR}$, which is consistent with what is seen in other galaxies. 
\par
4) Arp 220 is a unique and complicated system with ammonia seen in absorption for both metastable and non-metastable transitions. Thus a standard temperature analysis using only the ammonia metastable transitions breaks-down.  The relatively underpopulated (1,1) and (4,4) transitions in each nucleus indicate populated non-metastable states, contamination from intervening cool gas, or maser emission.  We consider the former possibility of populated non-metastable transitions as the most likely, but note that further observations of those transitions are needed for confirmation. 
\par
5)  The absorption profiles in these data are consistent with the absorption seen in HCN(4-3) and CS(7-6) observed with ALMA by \citet{2015ApJ...800...70S} for both nuclei, with stronger evidence in the western nucleus.  Depending on the systemic velocities of the two nuclei, this self-absorption could indicate a blue-shifted outflow extending to $\sim$400 km s$^{-1}$ in the western nucleus.
\par
6)  We detect the NH$_{3}$(9,9) transition in the western nucleus in spite of a non-detection with single-dish observations by \citet{2013ApJ...779...33M}, with the former non-detection likely due to it being diluted by the surrounding emission.
\par

\section{Acknowledgements}
 The National Radio Astronomy Observatory is a facility of the National Science Foundation operated under cooperative agreement by Associated Universities, Inc.  This paper makes use of the following ALMA data: ADS/JAO.ALMA\#2011.0.00175.S . ALMA is a partnership of ESO (representing its member states), NSF (USA) and NINS (Japan), together with NRC (Canada), NSC and ASIAA (Taiwan), and KASI (Republic of Korea), in cooperation with the Republic of Chile. The Joint ALMA Observatory is operated by ESO, AUI/NRAO and NAOJ.  This research made use of APLpy, an open-source plotting package for Python hosted at http://aplpy.github.com.  The authors thank Brent Groves and Yancy Shirley for insightful discussions concerning ammonia properties and LVG analysis.  We also thank Elizabeth Mills for insight concerning ammonia masers and column densities.  Finally, we thank the anonymous referee for exceptionally insightful and relevant comments leading to substantial improvement of this manuscript.

\pagebreak

\bibliographystyle{apj}
\bibliography{arp_220_paper}

\begin{thebibliography}{34}
\expandafter\ifx\csname natexlab\endcsname\relax\def\natexlab#1{#1}\fi

\bibitem[{{Araya} {et~al.}(2004){Araya}, {Baan}, \&
  {Hofner}}]{2004ApJS..154..541A}
{Araya}, E., {Baan}, W.~A., \& {Hofner}, P. 2004, \apjs, 154, 541

\bibitem[{{Barcos-Mu{\~n}oz} {et~al.}(2015){Barcos-Mu{\~n}oz}, {Leroy},
  {Evans}, {Privon}, {Armus}, {Condon}, {Mazzarella}, {Meier}, {Momjian},
  {Murphy}, {Ott}, {Reichardt}, {Sakamoto}, {Sanders}, {Schinnerer},
  {Stierwalt}, {Surace}, {Thompson}, \& {Walter}}]{2015ApJ...799...10B}
{Barcos-Mu{\~n}oz}, L., {et~al.} 2015, \apj, 799, 10

\bibitem[{{Bothwell} {et~al.}(2010){Bothwell}, {Chapman}, {Tacconi}, {Smail},
  {Ivison}, {Casey}, {Bertoldi}, {Beswick}, {Biggs}, {Blain}, {Cox}, {Genzel},
  {Greve}, {Kennicutt}, {Muxlow}, {Neri}, \& {Omont}}]{2010MNRAS.405..219B}
{Bothwell}, M.~S., {et~al.} 2010, \mnras, 405, 219

\bibitem[{{Cernicharo} {et~al.}(2006){Cernicharo}, {Pardo}, \&
  {Weiss}}]{2006ApJ...646L..49C}
{Cernicharo}, J., {Pardo}, J.~R., \& {Weiss}, A. 2006, \apjl, 646, L49

\bibitem[{{Danby} {et~al.}(1988){Danby}, {Flower}, {Valiron}, {Schilke}, \&
  {Walmsley}}]{1988MNRAS.235..229D}
{Danby}, G., {Flower}, D.~R., {Valiron}, P., {Schilke}, P., \& {Walmsley},
  C.~M. 1988, \mnras, 235, 229

\bibitem[{{Downes} \& {Eckart}(2007)}]{2007A&A...468L..57D}
{Downes}, D., \& {Eckart}, A. 2007, \aap, 468, L57

\bibitem[{{Hagiwara} \& {Edwards}(2015)}]{2015ApJ...815..124H}
{Hagiwara}, Y., \& {Edwards}, P.~G. 2015, \apj, 815, 124

\bibitem[{{Henkel} {et~al.}(1984){Henkel}, {Guesten}, {Downes}, {Thum},
  {Wilson}, \& {Biermann}}]{1984A&A...141L...1H}
{Henkel}, C., {Guesten}, R., {Downes}, D., {Thum}, C., {Wilson}, T.~L., \&
  {Biermann}, P. 1984, \aap, 141, L1

\bibitem[{{Henkel} {et~al.}(2005){Henkel}, {Peck}, {Tarchi}, {Nagar}, {Braatz},
  {Castangia}, \& {Moscadelli}}]{2005A&A...436...75H}
{Henkel}, C., {Peck}, A.~B., {Tarchi}, A., {Nagar}, N.~M., {Braatz}, J.~A.,
  {Castangia}, P., \& {Moscadelli}, L. 2005, \aap, 436, 75

\bibitem[{{Henkel} {et~al.}(1986){Henkel}, {Wouterloot}, \&
  {Bally}}]{1986A&A...155..193H}
{Henkel}, C., {Wouterloot}, J.~G.~A., \& {Bally}, J. 1986, \aap, 155, 193

\bibitem[{{Huettemeister} {et~al.}(1995){Huettemeister}, {Wilson},
  {Mauersberger}, {Lemme}, {Dahmen}, \& {Henkel}}]{1995A&A...294..667H}
{Huettemeister}, S., {Wilson}, T.~L., {Mauersberger}, R., {Lemme}, C.,
  {Dahmen}, G., \& {Henkel}, C. 1995, \aap, 294, 667

\bibitem[{{Mangum} {et~al.}(2013){Mangum}, {Darling}, {Henkel}, {Menten},
  {MacGregor}, {Svoboda}, \& {Schinnerer}}]{2013ApJ...779...33M}
{Mangum}, J.~G., {Darling}, J., {Henkel}, C., {Menten}, K.~M., {MacGregor}, M.,
  {Svoboda}, B.~E., \& {Schinnerer}, E. 2013, \apj, 779, 33

\bibitem[{{Mangum} \& {Shirley}(2015)}]{2015PASP..127..266M}
{Mangum}, J.~G., \& {Shirley}, Y.~L. 2015, \pasp, 127, 266

\bibitem[{{Mart{\'{\i}}n} {et~al.}(2011){Mart{\'{\i}}n}, {Krips},
  {Mart{\'{\i}}n-Pintado}, {Aalto}, {Zhao}, {Peck}, {Petitpas}, {Monje},
  {Greve}, \& {An}}]{2011A&A...527A..36M}
{Mart{\'{\i}}n}, S., {et~al.} 2011, \aap, 527, A36

\bibitem[{{McMullin} {et~al.}(2007){McMullin}, {Waters}, {Schiebel}, {Young},
  \& {Golap}}]{2007ASPC..376..127M}
{McMullin}, J.~P., {Waters}, B., {Schiebel}, D., {Young}, W., \& {Golap}, K.
  2007, in Astronomical Society of the Pacific Conference Series, Vol. 376,
  Astronomical Data Analysis Software and Systems XVI, ed. R.~A. {Shaw},
  F.~{Hill}, \& D.~J. {Bell}, 127

\bibitem[{{Mills} \& {Morris}(2013)}]{2013ApJ...772..105M}
{Mills}, E.~A.~C., \& {Morris}, M.~R. 2013, \apj, 772, 105

\bibitem[{{Mundell} {et~al.}(2001){Mundell}, {Ferruit}, \&
  {Pedlar}}]{2001ApJ...560..168M}
{Mundell}, C.~G., {Ferruit}, P., \& {Pedlar}, A. 2001, \apj, 560, 168

\bibitem[{{Nakai} {et~al.}(2002){Nakai}, {Sato}, \&
  {Yamauchi}}]{2002PASJ...54L..27N}
{Nakai}, N., {Sato}, N., \& {Yamauchi}, A. 2002, \pasj, 54, L27

\bibitem[{{Ott} {et~al.}(2011){Ott}, {Henkel}, {Braatz}, \&
  {Wei{\ss}}}]{2011ApJ...742...95O}
{Ott}, J., {Henkel}, C., {Braatz}, J.~A., \& {Wei{\ss}}, A. 2011, \apj, 742, 95

\bibitem[{{Rangwala} {et~al.}(2015){Rangwala}, {Maloney}, {Wilson}, {Glenn},
  {Kamenetzky}, \& {Spinoglio}}]{2015ApJ...806...17R}
{Rangwala}, N., {Maloney}, P.~R., {Wilson}, C.~D., {Glenn}, J., {Kamenetzky},
  J., \& {Spinoglio}, L. 2015, \apj, 806, 17

\bibitem[{{Remijan}(2010)}]{2010AAS...21547905R}
{Remijan}, A.~J. 2010, in Bulletin of the American Astronomical Society,
  Vol.~42, American Astronomical Society Meeting Abstracts \#215, 568

\bibitem[{{Sakamoto} {et~al.}(2009){Sakamoto}, {Aalto}, {Wilner}, {Black},
  {Conway}, {Costagliola}, {Peck}, {Spaans}, {Wang}, \&
  {Wiedner}}]{2009ApJ...700L.104S}
{Sakamoto}, K., {et~al.} 2009, \apjl, 700, L104

\bibitem[{{Salter} {et~al.}(2008){Salter}, {Ghosh}, {Catinella}, {Lebron},
  {Lerner}, {Minchin}, \& {Momjian}}]{2008AJ....136..389S}
{Salter}, C.~J., {Ghosh}, T., {Catinella}, B., {Lebron}, M., {Lerner}, M.~S.,
  {Minchin}, R., \& {Momjian}, E. 2008, \aj, 136, 389

\bibitem[{{Sanders} {et~al.}(2003){Sanders}, {Mazzarella}, {Kim}, {Surace}, \&
  {Soifer}}]{2003AJ....126.1607S}
{Sanders}, D.~B., {Mazzarella}, J.~M., {Kim}, D.-C., {Surace}, J.~A., \&
  {Soifer}, B.~T. 2003, \aj, 126, 1607

\bibitem[{{Sanders} \& {Mirabel}(1996)}]{1996ARA&A..34..749S}
{Sanders}, D.~B., \& {Mirabel}, I.~F. 1996, \araa, 34, 749

\bibitem[{{Scoville} {et~al.}(2015){Scoville}, {Sheth}, {Walter}, {Manohar},
  {Zschaechner}, {Yun}, {Koda}, {Sanders}, {Murchikova}, {Thompson},
  {Robertson}, {Genzel}, {Hernquist}, {Tacconi}, {Brown}, {Narayanan},
  {Hayward}, {Barnes}, {Kartaltepe}, {Davies}, {van der Werf}, \&
  {Fomalont}}]{2015ApJ...800...70S}
{Scoville}, N., {et~al.} 2015, \apj, 800, 70

\bibitem[{{Scoville} {et~al.}(2016){Scoville}, {Murchikova}, {Walter},
  {Vlahakis}, {Koda}, {Vanden Bout}, {Barnes}, {Hernquist}, {Sheth}, {Yun},
  {Sanders}, {Armus}, {Cox}, {Thompson}, {Robertson}, {Zschaechner}, {Tacconi},
  {Torrey}, {Hayward}, {Genzel}, {Hopkins}, {van der Werf}, \&
  {Decarli}}]{2016arXiv160509381S}
---. 2016, ArXiv e-prints

\bibitem[{{Scoville} {et~al.}(1986){Scoville}, {Sanders}, {Sargent}, {Soifer},
  {Scott}, \& {Lo}}]{1986ApJ...311L..47S}
{Scoville}, N.~Z., {Sanders}, D.~B., {Sargent}, A.~I., {Soifer}, B.~T.,
  {Scott}, S.~L., \& {Lo}, K.~Y. 1986, \apjl, 311, L47

\bibitem[{{Soifer} {et~al.}(1987){Soifer}, {Sanders}, {Madore}, {Neugebauer},
  {Danielson}, {Elias}, {Lonsdale}, \& {Rice}}]{1987ApJ...320..238S}
{Soifer}, B.~T., {Sanders}, D.~B., {Madore}, B.~F., {Neugebauer}, G.,
  {Danielson}, G.~E., {Elias}, J.~H., {Lonsdale}, C.~J., \& {Rice}, W.~L. 1987,
  \apj, 320, 238

\bibitem[{{Takano} {et~al.}(2005){Takano}, {Nakanishi}, {Nakai}, \&
  {Takano}}]{2005PASJ...57L..29T}
{Takano}, S., {Nakanishi}, K., {Nakai}, N., \& {Takano}, T. 2005, \pasj, 57,
  L29

\bibitem[{{Tunnard} {et~al.}(2015){Tunnard}, {Greve}, {Garcia-Burillo},
  {Graci{\'a} Carpio}, {Fischer}, {Fuente}, {Gonz{\'a}lez-Alfonso},
  {Hailey-Dunsheath}, {Neri}, {Sturm}, {Usero}, \&
  {Planesas}}]{2015ApJ...800...25T}
{Tunnard}, R., {et~al.} 2015, \apj, 800, 25

\bibitem[{{van der Tak} {et~al.}(2007){van der Tak}, {Black}, {Sch{\"o}ier},
  {Jansen}, \& {van Dishoeck}}]{2007A&A...468..627V}
{van der Tak}, F.~F.~S., {Black}, J.~H., {Sch{\"o}ier}, F.~L., {Jansen}, D.~J.,
  \& {van Dishoeck}, E.~F. 2007, \aap, 468, 627

\bibitem[{{Walmsley} \& {Ungerechts}(1983)}]{1983A&A...122..164W}
{Walmsley}, C.~M., \& {Ungerechts}, H. 1983, \aap, 122, 164

\bibitem[{{Wiggins} {et~al.}(2016){Wiggins}, {Migenes}, \&
  {Smidt}}]{2016ApJ...816...55W}
{Wiggins}, B.~K., {Migenes}, V., \& {Smidt}, J.~M. 2016, \apj, 816, 55

\end{thebibliography}

\end{document}